\documentclass[cameraready]{Interspeech}
\title{One-Step Token-to-Waveform Generation with MeanFlow in Latent Space}

\author[affiliation={1}]{Zheqi}{Dai}
\author[affiliation={2}]{Guangyan}{Zhang}
\author[affiliation={3}]{Zhen}{Ye}
\author[affiliation={2}]{Jingyu}{Li}
\author[affiliation={1}]{Haolin}{He}
\author[affiliation={1}]{Chunyat}{Wu}
\author[affiliation={4}, correspondingauthor]{Yiwen}{Guo}
\author[affiliation={1}, correspondingauthor]{Qiuqiang}{Kong}

\address{
    $^1$ The Chinese University of Hong Kong, Hong Kong SAR, China \\
    $^2$ LIGHTSPEED, Tencent, Hong Kong SAR, China \\
    $^3$ The Hong Kong University of Science and Technology, Hong Kong SAR, China \\
    $^4$ Independent Researcher
}

\email{zheqidai@link.cuhk.edu.hk, guoyiwen89@gmail.com, qqkong@ee.cuhk.edu.hk}


\newcommand{\blfootnote}[1]{%
  \begingroup
  \renewcommand\thefootnote{}\footnote{#1}%
  \addtocounter{footnote}{-1}%
  \endgroup
}
\usepackage{comment}
\usepackage{amsmath}
\usepackage{amssymb}
\usepackage{mathtools}
\usepackage{multirow}
\usepackage{bm}
\usepackage{placeins}
\usepackage{url}

\begin{document}

\maketitle

\blfootnote{Code\&Demo: \url{https://github.com/dzq84/meantok}}

\begin{abstract}
Neural audio codecs are central to modern LLM-based Text-to-Speech (TTS) and multimodal systems. As low-bitrate semantic codecs gain prominence, the Token-to-Waveform (Token2Wav) decoder becomes a bottleneck determining both perceptual quality and system efficiency. Conventional multi-step flow-matching decoders offer superior quality but suffer from high inference latency due to iterative sampling, creating a severe quality-speed trade-off. In this paper, we propose a novel Token2Wav architecture that overcomes this limitation by applying MeanFlow in a highly compressed latent space. By modeling the average velocity rather than the instantaneous velocity field, MeanFlow enables true one-step generation. Operating in the latent domain mitigates the memory and stability issues of waveform-level flows, yielding up to a 17$\times$ improvement in Real-Time Factor (RTF) compared to multi-step baselines with negligible quality degradation. Furthermore, we introduce refinement strategies that mitigate latent mismatch, including decoder-only fine-tuning with the MeanFlow generator frozen and end-to-end joint fine-tuning, improving fidelity without increasing inference-time cost. Code and demo are publicly available.
\end{abstract}

\section{Introduction}

Large language model (LLM)-based text-to-speech (TTS) systems~\cite{du2024cosyvoice2,li2025msr,ye2025llasa,guo2024fireredtts} increasingly adopt a discrete-token formulation: an upstream model predicts a sequence of speech tokens, and a downstream neural decoder converts these tokens into a waveform.
Neural audio codecs~\cite{siuzdak2024snac,li2025msr,zeghidour2021soundstream} provide a practical interface for such systems by discretizing speech into code sequences with a learned decoder.
In this setting, token design crucially affects the division of labor between the LLM and the decoder.
Acoustic tokens typically operate at higher bitrates and retain fine-grained signal details, making waveform reconstruction easier but increasing token rate and sequence length, which in turn raises modeling cost for LLMs~\cite{defossez2022high,zeghidour2021soundstream}
In contrast, semantic tokens~\cite{hsu2021hubert,liu2024semanticodec} are more compact and closer to linguistic content, improving generation efficiency and controllability.
However, as semantic tokenizations remove substantial acoustic detail, they shift the burden of recovering prosody, timbre, and naturalness to the Token-to-Waveform (Token2Wav) decoder.

This shift makes Token2Wav both critical and challenging: the decoder must synthesize high-fidelity speech from a heavily compressed, information-limited representation, while meeting the low-latency requirements of interactive and on-device applications.
Prior work on high-fidelity neural audio decoders~\cite{kumar2024high,siuzdak2024vocos} demonstrates that powerful discriminator-based training enables waveform reconstruction from compact latent representations.
Flow-matching-based generative decoders~\cite{lipman2023flow} have recently shown strong Token2Wav performance~\cite{du2024cosyvoice2,guo2024fireredtts}.
Despite their high quality, most flow-matching decoders rely on numerical integration of an ODE at inference time, requiring tens of network evaluations (NFEs) even with relatively coarse solvers.
This iterative sampling cost often dominates runtime and creates a persistent quality--latency trade-off.

MeanFlow~\cite{geng2025mean} reduces sampling cost by learning an \emph{average}-velocity field along the probability path, enabling one-step generation with a single network evaluation.
However, directly applying one-step flow models in waveform space remains difficult for high-fidelity conditional decoding.
Waveforms are extremely long sequences, making training memory-intensive and optimization unstable; moreover, a single large integration step amplifies modeling errors at raw-audio resolution.

In this paper, we propose a one-step Token2Wav architecture that applies MeanFlow in a \emph{compressed latent space}.
We first train a lightweight waveform variational autoencoder (VAE) that encodes speech into a low-dimensional latent sequence and deterministically decodes it back to waveform audio.
We then train a conditional 1D Diffusion Transformer (DiT) to perform MeanFlow-based \emph{token-to-latent} generation conditioned on semantic tokens and speaker embeddings.
At inference time, the latent sequence is generated in exactly one forward pass and converted to waveform by the VAE decoder, achieving true one-step Token2Wav synthesis.

A practical challenge of latent-space one-step generation is \emph{latent mismatch}: latents produced by the one-step generator may deviate from the VAE latents seen during VAE training, which can degrade waveform reconstruction.
To address this issue without increasing inference cost, we introduce refinement strategies that keep the inference pipeline unchanged (one generator pass plus one decoder pass): (i) \emph{decoder-only refinement} that freezes the MeanFlow generator and fine-tunes only the VAE decoder on generated latents, and (ii) \emph{end-to-end joint fine-tuning} that further backpropagates waveform-domain losses through the one-step latent sampling step.

\begin{figure*}[htbp]
    \centering
    \includegraphics[width=\textwidth,keepaspectratio]{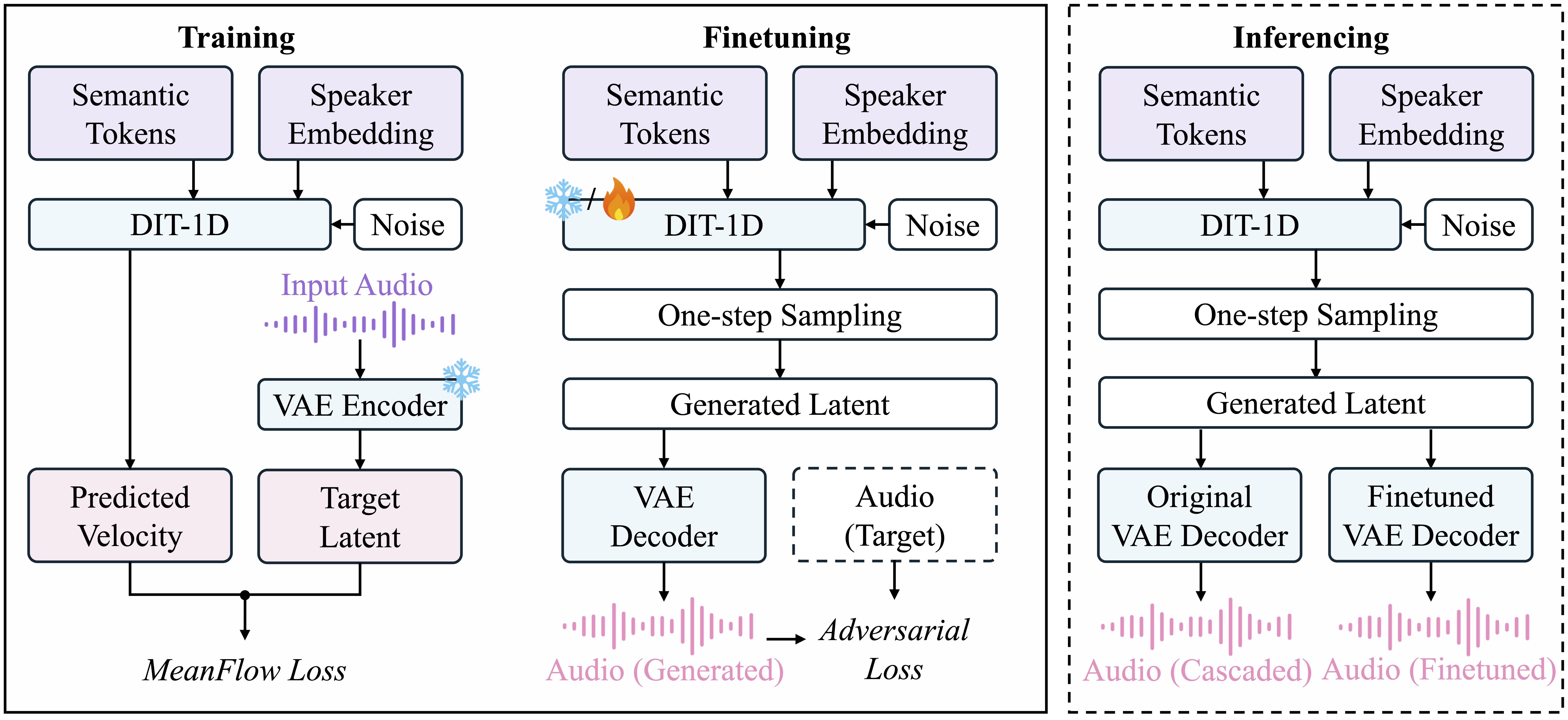}
    \caption{Overall framework. Left: MeanFlow training in latent space using VAE latents as targets. Middle: refinement during fine-tuning using waveform-domain losses on audio reconstructed from generated latents. Right: inference uses the same one-step sampling plus VAE decoding, with either the original or refined decoder.}
    \label{fig:overall-framework}
\end{figure*}

Figure~\ref{fig:overall-framework} summarizes the full pipeline and highlights. Results show that latent-space MeanFlow enables substantial latency reduction while maintaining competitive intelligibility and perceptual quality, achieving up to a 17$\times$ reduction in real-time factor (RTF) over a representative multi-step Token2Wav baseline under the same measurement protocol. Our main contributions are:
\begin{itemize}
    \item We propose a Token2Wav framework that performs MeanFlow-based \emph{one-step} conditional generation in a compressed latent space, avoiding the memory and stability issues of waveform-level flow models.
    \item We study the effect of latent dimensionality and model capacity on the quality--latency trade-off for one-step Token2Wav decoding, identifying effective operating points under tight latency constraints.
    \item We introduce refinement strategies---decoder-only refinement and end-to-end joint fine-tuning---to mitigate latent distribution mismatch, improving waveform fidelity without increasing inference-time cost.
\end{itemize}

\section{Method}

As shown in Figure~\ref{fig:overall-framework}, our Token2Wav decoder synthesizes waveform speech in two stages.
Given semantic tokens $\mathbf{s}=\{s_1,\dots,s_T\}$ and a speaker embedding $\mathbf{e}$, we denote the conditioning as $\mathbf{c}=(\mathbf{s},\mathbf{e})$ and aim to generate a waveform $\mathbf{x}\in\mathbb{R}^{L}$.
To achieve low latency, we (i) generate a compressed latent sequence in one step using a latent MeanFlow generator, and (ii) reconstruct waveform audio deterministically using a lightweight VAE decoder.
This design confines the one-step generative modeling to a short, low-dimensional sequence, improving both efficiency and stability compared to waveform-space one-step flows.

\subsection{Waveform VAE for Latent Representation}
\label{sec:vae}

Directly modeling conditional generation in waveform space is memory intensive and often unstable due to extremely long sequences.
We therefore operate in a compressed latent space using a lightweight waveform VAE, following the latent diffusion paradigm~\cite{rombach2022high}.
The VAE encoder $E_\phi$ maps a waveform segment $\mathbf{x}$ to a latent sequence
\begin{equation}
\mathbf{z}_{\text{data}} = E_\phi(\mathbf{x})\in\mathbb{R}^{T'\times D},
\label{eq:vae_enc}
\end{equation}
where $T'$ is the downsampled length and $D$ is the latent channel dimension per frame.
A deterministic VAE decoder $G_\psi$ reconstructs waveform audio as
\begin{equation}
\hat{\mathbf{x}} = G_\psi(\mathbf{z}).
\label{eq:vae_dec}
\end{equation}
At inference time, only $G_\psi$ is used as the waveform decoder, while $E_\phi$ is used only to obtain training latents $\mathbf{z}_{\text{data}}$ (Figure~\ref{fig:overall-framework}, left).

\subsection{Flow Matching and MeanFlow in Latent Space}
\label{sec:meanflow}

We train a conditional one-step generator in the latent space, implemented as a DiT-1D model and denoted by $f_\theta$ (Figure~\ref{fig:overall-framework}).
Compared to waveform-space generation, latent-space generation substantially reduces sequence length and dynamic range, making the large-step one-step update more stable and memory efficient.

Let $\boldsymbol{\epsilon}\sim\mathcal{N}(\mathbf{0},\mathbf{I})$ and define the rectified flow path~\cite{liu2023flow} in latent space:
\begin{equation}
\mathbf{z}_t = (1-t)\mathbf{z}_{\text{data}} + t\boldsymbol{\epsilon},\quad t\in[0,1],
\label{eq:rf_path}
\end{equation}
where we follow the reverse-time convention: $t=1$ corresponds to the Gaussian prior and $t=0$ corresponds to the data distribution.

\textbf{Conditional flow matching.}
Flow matching~\cite{lipman2023flow} learns a conditional velocity field $\mathbf{v}_\theta(\mathbf{z}_t,t,\mathbf{c})$ that matches the target velocity along the probability path.
For the rectified path in Eq.~\eqref{eq:rf_path}, the target instantaneous velocity is constant:
\begin{equation}
\mathbf{u}_{\text{rf}}(\mathbf{z}_t,t,\mathbf{c})
= \frac{d\mathbf{z}_t}{dt}
= \boldsymbol{\epsilon} - \mathbf{z}_{\text{data}}.
\label{eq:rf_target}
\end{equation}
Standard conditional flow matching minimizes
\begin{equation}
\mathcal{L}_{\text{CFM}}(\theta)=
\mathbb{E}_{t,\mathbf{x},\bm{\epsilon}}
\Big[\lVert\mathbf{v}_\theta(\mathbf{z}_t,t,\mathbf{c})-\mathbf{u}_{\text{rf}}\rVert_2^2\Big],
\label{eq:l_cfm}
\end{equation}
but requires multi-step ODE integration at inference time.

\textbf{MeanFlow for one-step generation.}
MeanFlow~\cite{geng2025mean} enables one-step sampling by predicting an \emph{average} velocity over an interval $[r,t]$:
\begin{equation}
\bar{\mathbf{u}}(\mathbf{z}_t;r,t,\mathbf{c})
\triangleq
\frac{1}{t-r}\int_{r}^{t}\mathbf{v}(\mathbf{z}_\tau,\tau,\mathbf{c})\,d\tau,\quad 0\le r<t\le 1.
\label{eq:mean_u_def}
\end{equation}
We use $f_\theta$ (DiT-1D) to predict $\bar{\mathbf{u}}$.
Following~\cite{geng2025mean}, we optimize the MeanFlow objective with stop-gradient and adaptive reweighting:
\begin{equation}
\mathcal{L}_{\text{MF}}(\theta)=
\mathbb{E}_{r,t,\mathbf{x},\bm{\epsilon}}
\Big[
w(r,t)\,
\big\lVert f_\theta(\mathbf{z}_t,r,t,\mathbf{c})-\operatorname{sg}(\bar{\mathbf{u}}_{\text{gt}})\big\rVert_2^2
\Big],
\label{eq:l_mf}
\end{equation}
where $\text{sg}(\cdot)$ denotes stop-gradient, $w(\cdot)$ is the adaptive weighting in~\cite{geng2025mean}, and $\bar{\mathbf{u}}_{\text{gt}}$ is constructed as in~\cite{geng2025mean}.
The required derivatives with respect to $t$ are computed efficiently using Jacobian-vector products (JVPs)~\cite{geng2025mean}.
When $r=t$, Eq.~\eqref{eq:l_mf} reduces to standard conditional flow matching.


\textbf{Conditioning.}
The generator is conditioned on semantic tokens and a speaker embedding.
Token embeddings, timestep embeddings, and speaker embeddings are fused into a conditioning sequence that modulates each DiT-1D block via adaptive layer normalization (adaLN-Zero)~\cite{peebles2023scalable}.

\textbf{One-step token-to-latent sampling.}
At inference time (Figure~\ref{fig:overall-framework}, right), we draw $\mathbf{z}_1\sim\mathcal{N}(\mathbf{0},\mathbf{I})$ and perform a single MeanFlow update:
\begin{equation}
\mathbf{z}_{\text{gen}} = \mathbf{z}_1 - f_\theta(\mathbf{z}_1,0,1,\mathbf{c}),
\label{eq:one_step_gen}
\end{equation}
followed by deterministic waveform reconstruction $\hat{\mathbf{x}} = G_\psi(\mathbf{z}_{\text{gen}})$.
This yields a fixed-cost inference pipeline consisting of one generator pass and one VAE decoder pass.

\subsection{Refinement for Latent Mismatch}
\label{sec:refine_joint}

Although latent-space one-step generation is efficient, the generated latent distribution may deviate from the VAE latent distribution induced by $E_\phi$, leading to reconstruction artifacts when decoding with $G_\psi$.
To reduce this \emph{latent mismatch} without increasing inference-time cost, we refine the model using waveform-domain losses while keeping the inference pipeline unchanged (Figure~\ref{fig:overall-framework}, middle and right).

We optimize the following waveform-domain objectives:
\begin{equation}
\mathcal{L}_{\text{ref}}
=
\mathcal{L}_{\text{MRSTFT}}
+
\lambda_{\text{adv}}\mathcal{L}_{\text{adv}}
+
\lambda_{\text{fm}}\mathcal{L}_{\text{fm}},
\label{eq:l_ref}
\end{equation}
where $\mathbf{z}_{\text{gen}}$ is obtained by Eq.~\eqref{eq:one_step_gen} and $\hat{\mathbf{x}}=G_\psi(\mathbf{z}_{\text{gen}})$.
$\mathcal{L}_{\text{MRSTFT}}$ is a multi-resolution STFT reconstruction loss, and $\mathcal{L}_{\text{adv}}$ and $\mathcal{L}_{\text{fm}}$ are adversarial and feature-matching losses computed using a multi-scale discriminator~\cite{defossez2022high}.
The discriminator is used only during training.

\textbf{Decoder-only refinement.}
We freeze the MeanFlow generator $f_\theta$ and update only the VAE decoder $G_\psi$ (and discriminator when used) to minimize Eq.~\eqref{eq:l_ref}.
This corresponds to adapting the decoder to the generator-induced latent distribution (Figure~\ref{fig:overall-framework}, middle).

\textbf{End-to-end joint fine-tuning.}
We update both $f_\theta$ and $G_\psi$ by backpropagating through the one-step update in Eq.~\eqref{eq:one_step_gen}.
The VAE encoder $E_\phi$ remains frozen.
This further reduces mismatch by improving both latent generation and waveform reconstruction, still keeping inference unchanged.

\section{Experiment}

\subsection{Experimental Setup and Metrics}
\label{sec:exp_setup}
\textbf{Datasets.}
We train all models on LibriTTS~\cite{zen2019libritts} and evaluate on the \textit{test-clean} subset of LibriSpeech~\cite{panayotov2015librispeech}.

\textbf{Tokenization and speaker conditioning.}
For fair comparison, we use the same semantic tokenization as the CosyVoice2 baseline~\cite{du2024cosyvoice2}.
Semantic tokens are extracted at 25~Hz using the CosyVoice2 tokenizer (single codebook; vocabulary size 6{,}561).
Speaker information is provided via a pretrained CAM++ speaker encoder~\cite{wang23ha_interspeech}, from which we extract a 192-dimensional speaker embedding.
Unless otherwise stated, all systems use identical conditioning signals.

\textbf{Model configurations.}
We vary the VAE latent dimensionality $D \in \{8,16,24\}$.
The MeanFlow latent generator is instantiated as a 1D DiT at two scales:
(i) 140M parameters (hidden size 768, 12 layers, 12 heads) and
(ii) 600M parameters (hidden size 1152, 28 layers, 16 heads).

\textbf{VAE architecture and training.}
The waveform VAE downsamples 24~kHz audio to a latent frame rate of 25~Hz to align with the semantic token rate.
We implement the encoder with an Oobleck-style strided convolution stack (strides $[2,4,4,6,5]$, total downsampling ratio 960) and use a deterministic decoder.
The VAE is trained on 2-second waveform chunks using a multi-resolution STFT reconstruction loss (FFT sizes from 32 to 2048), an adversarial hinge loss and a feature-matching loss with an EnCodec-style multi-scale discriminator~\cite{defossez2022high}, and a KL regularization term on the bottleneck.
We set $\lambda_{\text{adv}}{=}0.1$, $\lambda_{\text{fm}}{=}5.0$, and $\lambda_{\text{kl}}{=}10^{-4}$.
The discriminator is enabled after 1{,}000 warmup steps.
\begin{table*}[!t]
\centering
\caption{Main results on LibriSpeech test-clean. RTF is end-to-end (generator + decoder); stage-wise breakdown for our best system: DiT 0.0016 + VAE 0.0030. MOS is reported as mean$\pm$95\% CI.}
\label{tab:main_results}
\setlength{\tabcolsep}{5pt}
\small
\begin{tabular}{>{\raggedright\arraybackslash}p{0.38\textwidth} c c c c c c}
\toprule
\textbf{System} & \textbf{Dim} & \textbf{WER(\%)} $\downarrow$ & \textbf{SpkSim} $\uparrow$ & \textbf{UTMOS} $\uparrow$ & \textbf{MOS} $\uparrow$ & \textbf{RTF} $\downarrow$ \\
\midrule
CosyVoice2 Token2Wav (10-step)~\cite{du2024cosyvoice2} & -- & 3.18 & 0.940 & 3.76 & 4.05$\pm$0.03 & 0.0775 \\
VAE reconstruction (oracle latent) & 24 & 2.14 & 0.966 & 3.67 & 4.10$\pm$0.04 & -- \\
\midrule
Latent MeanFlow + VAE (Joint-FT) & 24 & 3.41 & 0.932 & 3.64 & 3.85$\pm$0.03 & 0.0046 \\
Latent MeanFlow + VAE (Joint-FT) & 16 & 3.62 & 0.927 & 3.56 & 3.72$\pm$0.03 & 0.0046 \\
\bottomrule
\end{tabular}
\end{table*}

\textbf{MeanFlow generator training.}
The latent MeanFlow DiT is trained on 5-second segments to better capture longer-range semantic--acoustic dependencies.
We sample $(r,t)$ using the logit-normal scheme in~\cite{evans2025stable} with $\mu=-0.4$ and $\sigma=1.0$, and follow~\cite{geng2025mean} for the MeanFlow objective, including JVP-based derivative computation and adaptive loss reweighting.

\textbf{Baselines.}
Our primary baseline is the Token2Wav module of CosyVoice2~\cite{du2024cosyvoice2}, which performs \emph{token-to-mel} conditional flow matching with multi-step ODE sampling, followed by a pretrained neural vocoder to synthesize waveform audio from the predicted mel-spectrogram.
We evaluate CosyVoice2 with its default 10-step Euler sampler.
We also report a \emph{VAE-only} reconstruction upper bound by encoding ground-truth waveforms into VAE latents and decoding them back using the VAE decoder.

\textbf{Metrics.}
Since LibriSpeech references are natively 16~kHz while our models synthesize 24~kHz audio, all generated waveforms are resampled to 16~kHz before computing intrusive metrics.
We report:
(1) \textbf{WER} (\%), obtained by a fine-tuned HuBERT-Large ASR model~\cite{hsu2021hubert};
(2) \textbf{SpkSim}, cosine similarity computed from a fine-tuned WavLM-Large~\cite{chen2022wavlm} speaker verification model;
(3) \textbf{UTMOS}~\cite{saeki2022utmos}; and
(4) \textbf{RTF}, measured with batch size 1 on a single GPU.

\textbf{RTF measurement protocol.}
We measure RTF using FP16 inference with batch size 1 on an NVIDIA H20 GPU.
For our method, RTF includes both the one-step latent generator $f_\theta$ and the VAE decoder $G_\psi$.
For the CosyVoice2 baseline, RTF includes the 10-step token-to-mel flow sampling \emph{and} the pretrained vocoder inference.
All models synthesize 24~kHz audio; resampling to 16~kHz for metric computation is excluded from RTF.
Runtime is measured with CUDA synchronization and reported as the average over evaluation utterances.

\textbf{Subjective evaluation.}
We conduct a MOS test on 50 randomly selected utterances rated by 20 listeners on a 1--5 scale, and report MOS with 95\% confidence intervals.

\subsection{Main Results}
Table~\ref{tab:main_results} compares our proposed one-step Token2Wav decoder with the CosyVoice2 Token2Wav baseline~\cite{du2024cosyvoice2}, which performs token-to-mel conditional flow matching with 10-step Euler sampling followed by a pretrained vocoder.
Our best configuration, \textbf{Latent MeanFlow + VAE (Joint-FT)} with $D{=}24$ and a 140M DiT, achieves an RTF of 0.0046, a 17$\times$ end-to-end speedup over the multi-step baseline (RTF 0.0775).
Despite using only a single evaluation of the latent generator, it maintains competitive intelligibility and speaker similarity (WER 3.41\%, SpkSim 0.932), and reaches UTMOS 3.64 with MOS 3.85, approaching the baseline (UTMOS 3.76, MOS 4.05).
Finally, the \textbf{VAE reconstruction (oracle latent)} upper bound indicates that most remaining degradation is attributable to latent generation rather than waveform decoding, suggesting further headroom from improving the token-to-latent generator.
The stage-wise RTF breakdown in Table~\ref{tab:main_results} (caption) shows that our speed advantage comes from replacing iterative flow sampling with a single latent DiT evaluation, while the VAE decoder contributes a smaller but non-negligible fraction of the total cost.

\subsection{Ablation Studies}
\label{sec:ablations}

Table~\ref{tab:ablations} consolidates three ablation experiments under the \textbf{Latent MeanFlow + VAE} framework with a 140M DiT unless noted otherwise.

\textbf{Latent dimensionality} (Table~\ref{tab:ablations}, top).
Increasing $D$ consistently improves all quality metrics: from $D{=}8$ to $D{=}24$, WER decreases from 4.82\% to 3.41\%, SpkSim improves from 0.909 to 0.932, and MOS increases from 3.45 to 3.85.
We observe similar RTF across different $D$ because the one-step generator dominates runtime and the decoder cost scales mildly with $D$ at these operating points.
Overall, $D{=}16$--$24$ provides a practical quality--compression trade-off, and we use $D{=}24$ as our default setting in subsequent experiments.

\textbf{Model capacity} (Table~\ref{tab:ablations}, middle).
The 140M model matches or slightly outperforms the 600M variant on perceptual metrics (UTMOS 3.64 vs.\ 3.57; MOS 3.85 vs.\ 3.78) while also being faster (RTF 0.0047 vs.\ 0.0075).
These results suggest that simply scaling DiT size does not necessarily improve one-step decoding quality under the current data and training recipe.
One-step generation is sensitive to large-step integration error: larger models may overfit to local training dynamics and produce less robust average-velocity estimates under the single-step update.
We conjecture that larger-capacity models could become beneficial with longer training, stronger regularization, or improved conditioning tailored for one-step sampling.

\textbf{Refinement strategies} (Table~\ref{tab:ablations}, bottom).
A key challenge in combining a one-step latent generator with a VAE waveform decoder is latent distribution mismatch: generated latents may deviate from the VAE training latents, leading to reconstruction artifacts.
Directly composing the pretrained decoder with the one-step generator (\textbf{No-FT}) yields degraded perceptual quality (UTMOS 3.11, MOS 3.35), confirming the mismatch issue.
\textbf{Decoder-FT} improves quality (UTMOS 3.43, MOS 3.70) by adapting $G_\psi$ to the generator's latent distribution while keeping $f_\theta$ frozen.
\textbf{Joint-FT} achieves the best overall results (WER 3.41\%, UTMOS 3.64, MOS 3.85) by backpropagating waveform-domain objectives through the one-step latent sampling step, improving both latent generation and waveform reconstruction without increasing inference-time cost.
\begin{table}[!t]
\centering
\caption{Ablation studies (140M DiT, Joint-FT unless noted). MOS: mean$\pm$95\% CI.}
\label{tab:ablations}
\setlength{\tabcolsep}{3.5pt}
\small
\begin{tabular}{l c c c c}
\toprule
\textbf{Config} & \textbf{WER(\%)}$\downarrow$ & \textbf{SpkSim}$\uparrow$ & \textbf{UTMOS}$\uparrow$ & \textbf{MOS}$\uparrow$ \\
\midrule
\multicolumn{5}{l}{\emph{Latent dimensionality ($D$)}} \\
\quad $D{=}8$  & 4.82 & 0.909 & 3.47 & 3.45$\pm$0.03 \\
\quad $D{=}16$ & 3.62 & 0.927 & 3.56 & 3.72$\pm$0.03 \\
\quad $D{=}24$ & \textbf{3.41} & \textbf{0.932} & \textbf{3.64} & \textbf{3.85$\pm$0.04} \\
\midrule
\multicolumn{5}{l}{\emph{Model capacity ($D{=}24$)}} \\
\quad 140M & \textbf{3.41} & \textbf{0.932} & \textbf{3.64} & \textbf{3.85$\pm$0.03} \\
\quad 600M & 3.44 & 0.930 & 3.57 & 3.78$\pm$0.03 \\
\midrule
\multicolumn{5}{l}{\emph{Refinement strategy ($D{=}24$)}} \\
\quad No-FT      & 3.52 & 0.931 & 3.11 & 3.35$\pm$0.03 \\
\quad Decoder-FT  & 3.43 & 0.931 & 3.43 & 3.70$\pm$0.03 \\
\quad Joint-FT   & \textbf{3.41} & \textbf{0.932} & \textbf{3.64} & \textbf{3.85$\pm$0.04} \\
\bottomrule
\end{tabular}
\end{table}

\FloatBarrier
\section{Conclusion}

We presented a one-step Token2Wav decoder that applies MeanFlow in a highly compressed latent space to eliminate the iterative sampling overhead of flow-matching decoders.
The proposed system combines a latent MeanFlow generator (DiT-1D) that performs token-to-latent generation in a single network evaluation with a deterministic VAE decoder for latent-to-waveform reconstruction, resulting in a fixed-cost and low-latency inference pipeline.
To address latent mismatch between generated latents and the VAE training distribution, we introduced refinement strategies that optimize waveform-domain objectives while keeping inference unchanged.
Experiments on LibriSpeech~\cite{panayotov2015librispeech} test-clean demonstrate that our best configuration (140M DiT-1D, $D{=}24$, joint fine-tuning) achieves competitive intelligibility (WER 3.41\%) and perceptual quality, while delivering up to a 17$\times$ RTF speedup over a representative multi-step Token2Wav baseline under the same measurement protocol, suggesting that latent-space MeanFlow is a practical approach for real-time and on-device Token2Wav decoding under tight latency constraints.

\section{Generative AI Use Disclosure}

Generative AI tools were used for manuscript editing and polishing. All authors are responsible and accountable for the work and content of this paper.

\bibliographystyle{IEEEtran}
\bibliography{mybib}

\end{document}